\documentclass[aip, apl, amsmath, amssymb, longbibliography]{revtex4-1}
\usepackage[utf8]{inputenc}
\usepackage[british]{babel}
\usepackage{graphicx}
\usepackage{amsmath}
\usepackage{amsfonts}
\usepackage{color}

\usepackage{hyperref}
\begin{document}
\title{Electron Bessel beam diffraction for precise and accurate nanoscale strain mapping}
\date{\today}

\author{Giulio Guzzinati}
\email{giulio.guzzinati@uantwerpen.be}
\affiliation{EMAT, University of Antwerp, Groenenborgerlaan 171, 2020 Antwerp, Belgium}

\author{Wannes Ghielens}
\affiliation{Adrem Data Lab, University of Antwerp, Middelheimlaan 1, 2020 Antwerp, Belgium}

\author{Christoph Mahr}
\affiliation{Institute of Solid State Physics, University of Bremen, Otto-Hahn-Allee 1, D-28359 Bremen, Germany}
\affiliation{MAPEX Center for Materials and Processes, University of Bremen, Bibliothekstr. 1, D-28359 Bremen, Germany}

\author{Armand Béché}
\affiliation{EMAT, University of Antwerp, Groenenborgerlaan 171, 2020 Antwerp, Belgium}

\author{Andreas Rosenauer}
\affiliation{Institute of Solid State Physics, University of Bremen, Otto-Hahn-Allee 1, D-28359 Bremen, Germany}
\affiliation{MAPEX Center for Materials and Processes, University of Bremen, Bibliothekstr. 1, D-28359 Bremen, Germany}

\author{Toon Calders}
\affiliation{Adrem Data Lab, University of Antwerp, Middelheimlaan 1, 2020 Antwerp, Belgium}

\author{Jo Verbeeck}
\affiliation{EMAT, University of Antwerp, Groenenborgerlaan 171, 2020 Antwerp, Belgium}

\begin{abstract}
Strain has a strong effect on the properties of materials and the performance of electronic devices. Their ever shrinking size translates into a constant demand for accurate and precise measurement methods with very high spatial resolution. In this regard, transmission electron microscopes are key instruments thanks to their ability to map strain with sub-nanometer resolution.
Here we present a novel method to measure strain at the nanometer scale based on the diffraction of electron Bessel beams. We demonstrate that our method offers a strain sensitivity better than $2.5 \cdot 10^{-4}$ and an accuracy of $1.5 \cdot 10^{-3}$, competing with, or outperforming, the best existing methods with a simple and easy to use experimental setup.

\end{abstract}

\maketitle

Strain in materials is of extreme importance in a wide variety of technological applications and particularly in nanoelectronics where, besides impacting the devices' lifetime and performance, it is used deliberately for engineering the electronic transport properties \cite{Cressler2008,Chu2009}. The ever shrinking size of electronic devices translates into an increasing demand for accurate and precise strain measurement methods with a very high spatial resolution. 

When engineering nanomaterials, characterising nanostructures or current generation nanoelectronic devices, spatial resolution becomes paramount. Techniques such as X-Ray diffraction or $\mu$-Raman spectroscopy are quite easily accessible methods and  offer very good strain sensitivity (respectively, $\sigma = 1 \times 10^{-5}$ \cite{Auth2008} and $\sigma = 1 \times 10^{-4}$ \cite{Senez2003}) but a spatial resolution too low to investigate strain within a single device \cite{Xiong2014}.
Transmission electron microscopy (TEM) offers strain measurement with the highest spatial resolution and many methods have been developed or adapted to this purpose.
Strain can be measured through convergent beam electron diffraction (CBED), high resolution conventional and scanning TEM imaging (HRTEM and HRSTEM) including Moiré fringe analysis (Moiré fringes appear when using a HRSTEM setup to scan at a lower magnification, due to the low-frequency sampling of the crystal lattice)  as well as nano beam electron diffraction (NBED) which can also be performed in conjunction with precession electron diffraction (N-PED)\cite{Beche2013b, Kim2013, Ishizuka2017, Pofelski2018, Cooper2016}.
Of these techniques, the best accuracy and precision are offered by nano beam precession electron diffraction with a spatial resolution better than $1\,nm$, a strain sensitivity of $\sigma = 2 \times 10^{-4}$ and an accuracy of $ \Delta = 1 \times 10^{-3}$, though it requires additional specialised hardware \cite{Rouviere2013,Mahr2015}.
Conventional nano beam electron diffraction is also a highly precise ($\sigma = 6\times 10^{-4} $) and accurate ($\Delta = 2\times 10^{-3} $) technique, though it is not on par with N-PED and suffers the same limits in term of resolution.

Here we propose a new diffraction-based strain measurement protocol which can be applied on unmodified transmission electron microscopes by just replacing an aperture of the illumination system. It offers a simplified setup and performances approaching those of N-PED and clearly outperforming NBED, while providing excellent spatial resolution.

\begin{figure}[ht]
 \includegraphics[width=\columnwidth]{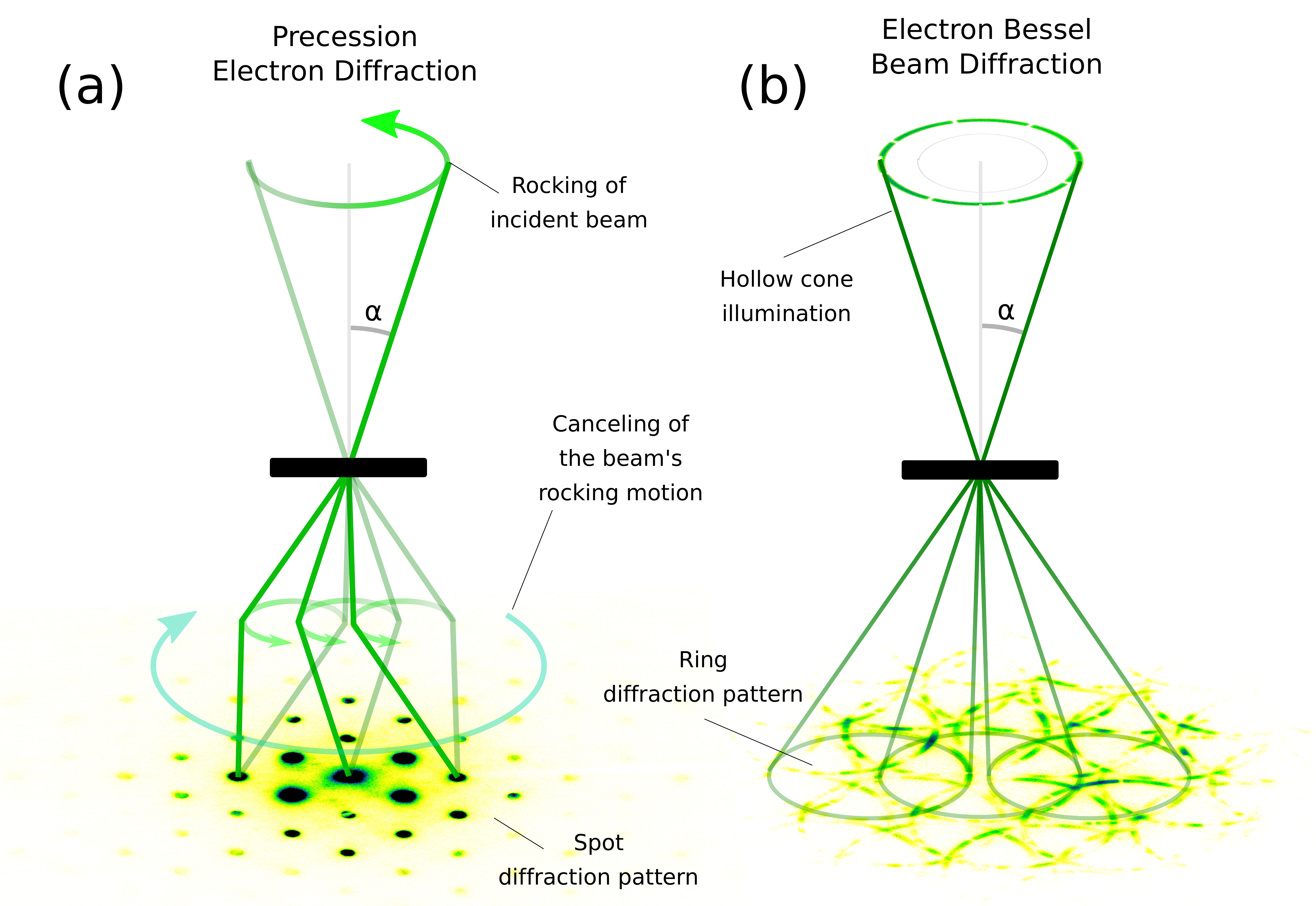}
 \caption{Schematic representation of our proposed setup in relation to PED. (a) In precession electron diffraction a low-convergence electron beam impinges on the sample under a fixed polar angle $\alpha$, while the azimuthal angle is rotated using deflectors. In order to still obtain a diffraction pattern formed of spots, another set of coils is used to cancel this rotation. The diffraction pattern is averaged over the different configurations, reducing dynamical effects. (b) In our proposal, we use an annular aperture to realise a hollow-cone illumination with semi-convergence angle $\alpha$, so that the rays from all the directions of incidence are simultaneously present. Due to this, the diffraction pattern is now constituted of rings making it significantly harder to analyse. \label{fig:schematic}}
\end{figure}

A fundamental problem with electron diffraction experiments is dynamical scattering. Due to the very strong interaction between the electron beam and the sample, the electrons are scattered many times within the specimen leading to several unwanted effects, from nonlinearities in the intensity of the diffracted beams, to a non-homogeneous shape of the diffracted discs, complicating significantly the extraction of structural information \cite{Muller2012b}.
Precession electron diffraction mitigates these problems by varying the beam's incident direction \cite{Eggeman2012}.
By entering the sample under a shallow angle (below 2 degrees, but below 0.5 degrees for strain applications) with respect to one of the main crystallographic directions, the electrons are scattered more weakly and therefore multiple scattering becomes less common. While keeping the angle of incidence constant, the direction of incidence is varied azimuthally so that the pattern is averaged over the different configurations. This leads to quasi-kinematical diffraction patterns which are less sensitive to local sample variation (thickness, bending...) and are thus easier to interpret \cite{Eggeman2012}.
This is typically achieved by using dedicated control hardware to rock the incident beam and then cancel this rocking after the interaction with the sample by using the microscope’s deflecting coils, so that the patterns for all different incident directions superimpose and are averaged, as schematically shown in Figure \ref{fig:schematic}a.
While the coils used are already part of the microscope, it's generally not possible or practical for the user to freely program them, making the purchase (or construction) of dedicated hardware necessary in most cases.
This, along with the alignment procedures needed to get the two rocking processes to compensate as well as possible, has kept this powerful method from gaining widespread adoption.

\begin{figure}[ht]
    \includegraphics[width=\columnwidth]{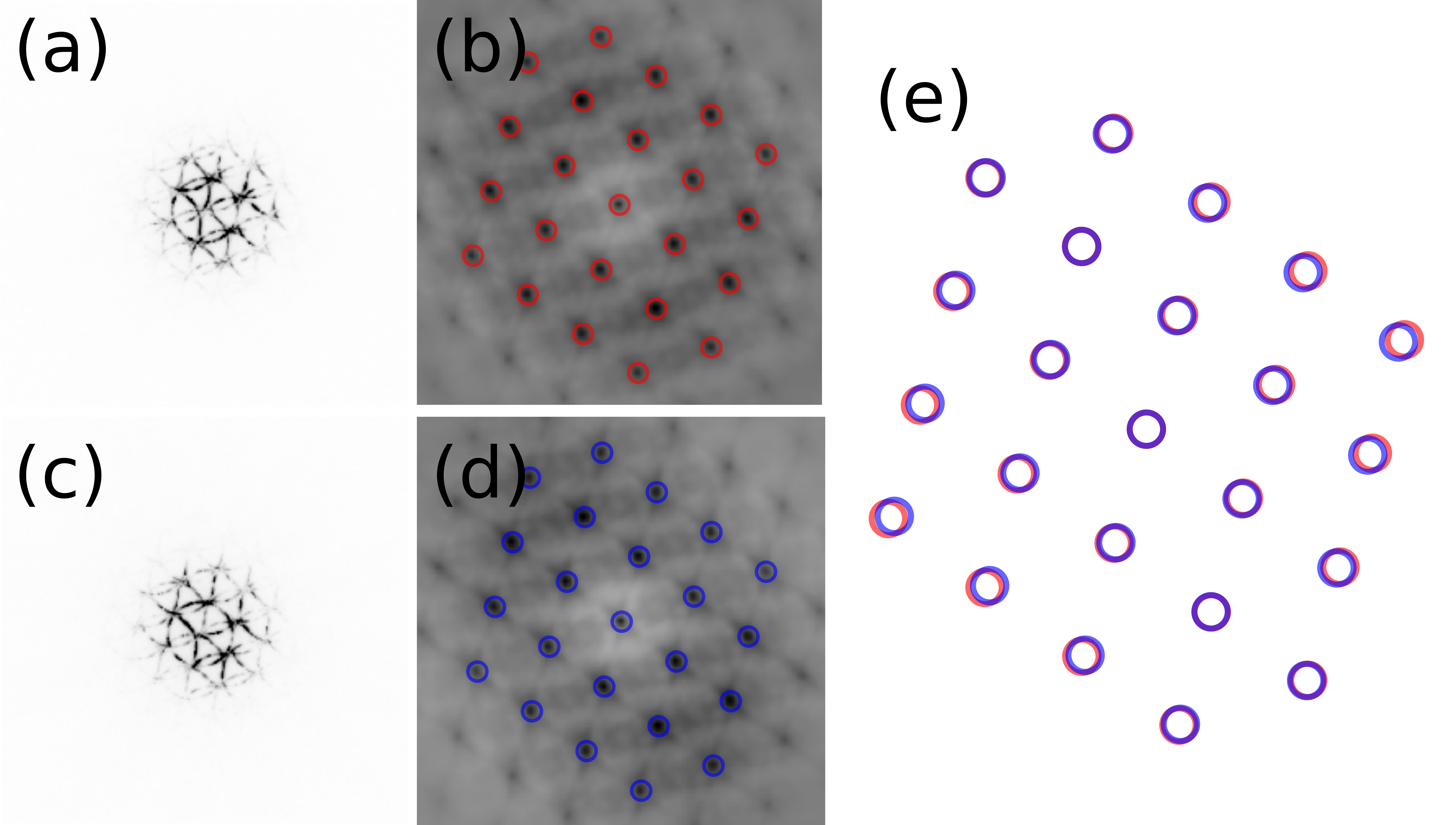}
    \caption{Ring diffraction patterns and analysis. (a) Ring diffraction pattern from the unstrained silicon substrate. (b) Autocorrelation function of the pattern, with marked positions for the maxima. (c) Ring diffraction pattern from a SiGe layer, significantly strained with respect to the substrate. (d) Autocorrelation function for the pattern in (c), also with marked peaks. (e) The extracted peaks differ visibly when overlayed. The strain is calculated by computing the affine transformation of the unstrained pattern that best allows to overlap it to the strained one. \label{fig:patterns}}
\end{figure}

Our proposal is inspired by precession, but rather than using different incidence directions in a sequential fashion, we realise a hollow-cone illumination where the rays from the different directions are present simultaneously.
This creates a different type of diffraction pattern where each spot is replaced by a ring which, for typical precession angles, overlap creating a more complex diffraction pattern (see figure \ref{fig:patterns}), which required the development of a dedicated analysis protocol.
In the following we will test this method through computer simulations, then demonstrate it experimentally and show that it offers performance close to those of N-PED without the extra equipment, making this technique potentially useful to a much wider audience of electron microscopists and materials scientists.


The core idea behind this work was to implement a ``parallel precession'' experiment, which would have had a wide number of experimental advantages. 
To understand how our proposed setup is equivalent to precession, one can think of the ring diffraction pattern as a convergent beam diffraction pattern where one has removed the central part of the diffracted disks, as in figure \ref{fig:patterns}a and \ref{fig:patterns}c.
In a diffraction pattern acquired with a parallel incident beam, the intensity of the diffracted spots depends on the exact direction of incidence and is very sensitive to even a slight tilt, in what is called the "rocking curve".
In CBED the diffraction pattern is formed of disks possessing non-homogeneous intensity. Each point in the disk formed by the transmitted beam corresponds to a different direction of incidence, and is associated to the corresponding diffracted points in the same position of the diffracted disks. The intensities of these points are identical to the intensities of the beams in a parallel-beam diffraction pattern with the same tilted direction of incidence \cite{Morniroli2002,Morniroli2008}. 
Keeping this in mind, it is clear that our approach yields exactly the same information as precession, as long as the rings do not overlap, i.e. as long as the semi-convergence angle $\alpha$ is lower than the Bragg angle for the sample under investigation.
For overlapping rings this is still largely valid, with the exception of the narrow overlap regions, where multi-beam effects are possible and further complicate the interpretation \cite{Morniroli2002}.

Similar diffraction patterns have also been realised before with a PED system by not ``untilting'' the beam in the projection system \cite{RouvierePatent2011,Rouviere2013}, and have been proposed as a way to improve precision by increasing the area of each spot \cite{RouvierePatent2011}.
Still, even in the simplest case of a non-overlapping ring pattern the reliable extraction of the ring positions has proven to be challenging \cite{Mahr2015}.

To explore the potential of our idea, we tested it using simulated series of diffraction patterns. We performed multislice simulations with the STEMsim software \cite{Rosenauer2008,VanDyck2009,Prestat2013, Krause2013} using the same model structure as in Ref. \cite{Mahr2015,Schowalter2009,Muraki1992,Grieb2017}. This structure is a Si-sample containing two embedded SiGe layers with different Ge-concentrations, of 38\% and 31\%, respectively, for details see Supplementary Information.

\begin{figure}
    \centering
    \includegraphics[width=\columnwidth]{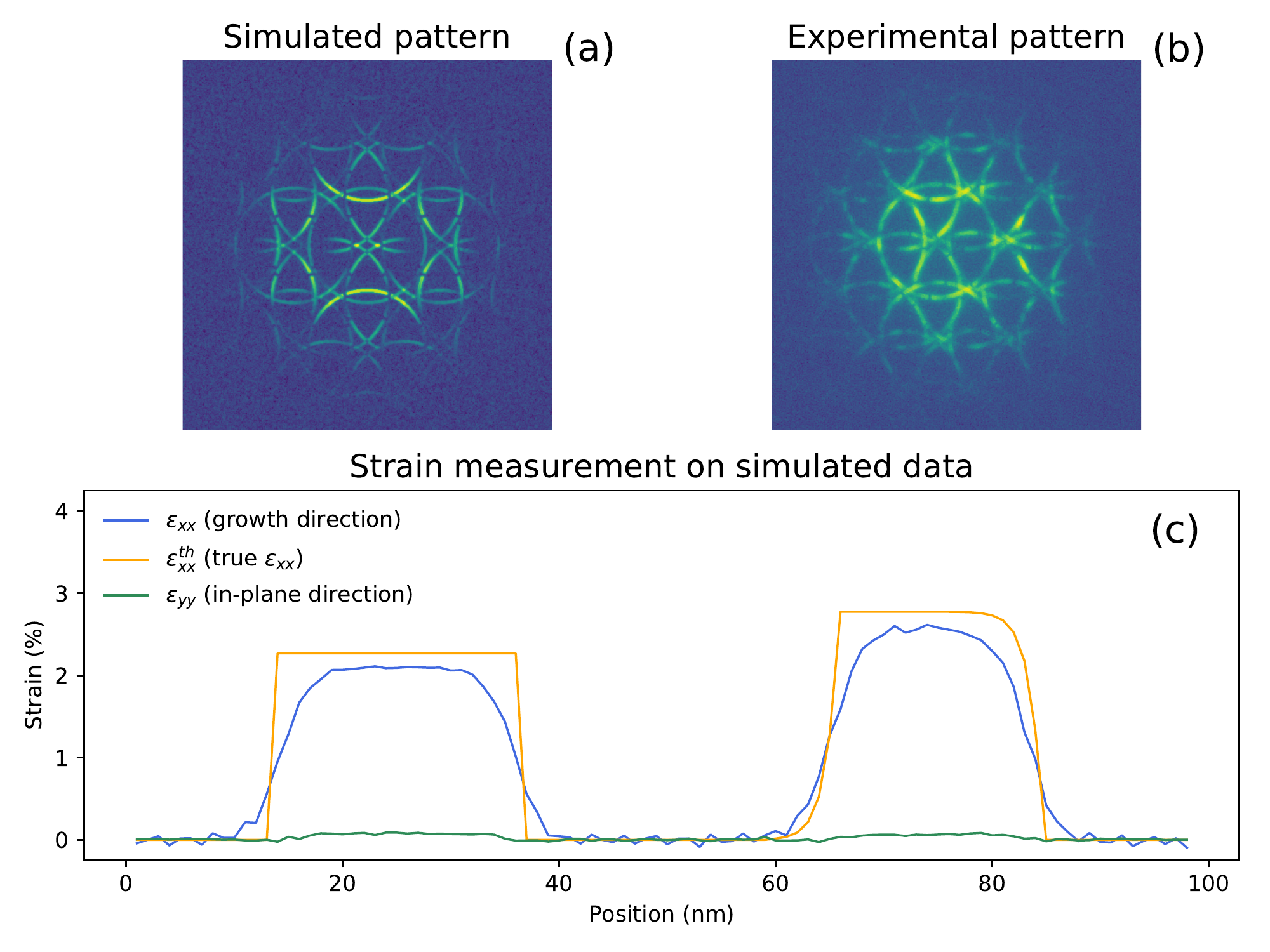}
    \caption{Simulated strain measurement. Bessel beam diffraction patterns have been simulated (a)  to reproduce experimental parameters and conditions (b). These patterns have been analyzed with our in house developed algorithm to assess the ability of this approach to effectively detect strain (c). The results show that the approach is accurate, though noisy.}
    \label{fig:simulations}
\end{figure}

While the resulting pattern is quite rich in details, it's worth remembering that strain measurement is a relatively simple problem. From all the wealth of details contained in this pattern we seek only to measure the three in-plane strain components: the two normal strain components $\epsilon_{xx}$ and $\epsilon_{yy}$ and the shear strain  $\epsilon_{xy}$, which are linked to the spacing between the Bragg reflections.

To this purpose we based our analysis on the autocorrelation function. For a two-dimensional signal $f(\mathbf{q}_\perp)$ its autocorrelation function $\mathcal{A}_{f}(\mathbf{q^\prime_\perp})$ is:
\begin{align}
    \mathcal{A}_{f}(\mathbf{q^\prime_\perp}) &=
    \iint f(\mathbf{q_\perp}+\mathbf{q^\prime_\perp})f^\star (\mathbf{q_\perp})\, \mathrm{d}\mathbf{q_\perp} \label{eq:autocorr} \\
     &= \iint f(\mathbf{q_\perp})f^\star(\mathbf{q_\perp}-\mathbf{q^\prime_\perp})\, \mathrm{d}\mathbf{q_\perp} \nonumber
\end{align}
or, using the Wiener-Kinchin theorem:
\begin{align}
    \mathcal{A}_{f} (\mathbf{q^\prime_\perp}) = \mathcal{F}^{-1}\left( \left| \mathcal{F} \left( f(\mathbf{q_\perp}) \right) \right|^2  
    \right) \label{eq:wiener}
\end{align}

The autocorrelation function is essentially a measure of self-similarity, that is of similarity between different parts of the same signal (here, a diffraction pattern).
For a signal containing a certain periodicity the autocorrelation function will display peaks corresponding to that same period, as shown in figure \ref{fig:patterns}b and \ref{fig:patterns}d. This means we get a peak when the shift between the diffraction pattern and its copy is such that the (0,0,0) ring in the copy overlaps a Bragg diffracted ring, e.g. the (0,0,2) in the original pattern, and the opposite (0,0,-2) diffracted ring overlaps with the (0,0,0) ring of the original pattern.
This superposition and integration of the ring intensities, in our approach, has a role riminiscent of the de-rocking procedure in conventional precession, averaging over the different parts of the rocking curve.
It is also worth noting that in every circumstance the autocorrelation function is always centered (i.e. the overlap between two copies of the pattern is highest with no shift) and is centrosymmetric (as shown in equation \ref{eq:autocorr}), which simplifies its analysis. 

The peaks appear to be small and sitting over a large background, which is due to the fact that even when the rings of the copies of the pattern are not perfectly aligned, the overlap is still significant.
To remove its effect we normalise the contrast radially by fitting the background's strength to a polynomial, then locate the peaks and fit their shape to a high-order polynomial surface, from which the  peak's position is extracted with sub-pixel precision. Once these positions have been extracted, the problem is reduced to finding the affine transform that best overlaps them with those from a chosen reference.
This is done by minimising:
\begin{equation}
\sum_\mathrm{i} \left| \left| 
\left[ \begin{array}{c}  \overline{q}^\mathrm{i}_{x} \\ \overline{q}^\mathrm{i}_{y} \end{array} \right]
- \left[ \begin{array}{cc} c_{xx} & c_{xy} \\ c_{yx} & c_{yy} \end{array} \right]
\cdot\left[ \begin{array}{c}q^\mathrm{i}_{x} \\ q^\mathrm{i}_{y} \end{array} \right]
\right| \right|^2
\end{equation}
where $\mathbf{q}^\mathrm{i} = (q^\mathrm{i}_{x},q^\mathrm{i}_{y})$ are the detected peak positions in the diffraction patterns, and $\overline{\mathbf{q}}^\mathrm{i} = (\overline{q}^\mathrm{i}_{x},\overline{q}^\mathrm{i}_{y})$ are the ones for the reference pattern. This allows to immediately obtain the strain components:
\begin{align}
    \varepsilon_{xx} &= c_{xx}-1,& \varepsilon_{yy} &= c_{yy}-1, \\
    \varepsilon_{xy} &= (c_{xy}+c_{yx})/2,& \omega &= (c_{xy}-c_{yx})/2. \nonumber
\end{align}
    
Obviously the reference pattern also contains noise and any error in its analysis impacts the performance of the technique, since the fitting would determine the transformation necessary to adapt each diffraction pattern to an inaccurate reference. To reduce this effect we have averaged the positions extracted from several patterns, recorded in a region which is supposed free of strain. The resulting code is freely available under a GPLv3 Licence \cite{GhielensCode,DataDrop}.

\begin{figure}[ht]
    \includegraphics[width=\columnwidth]{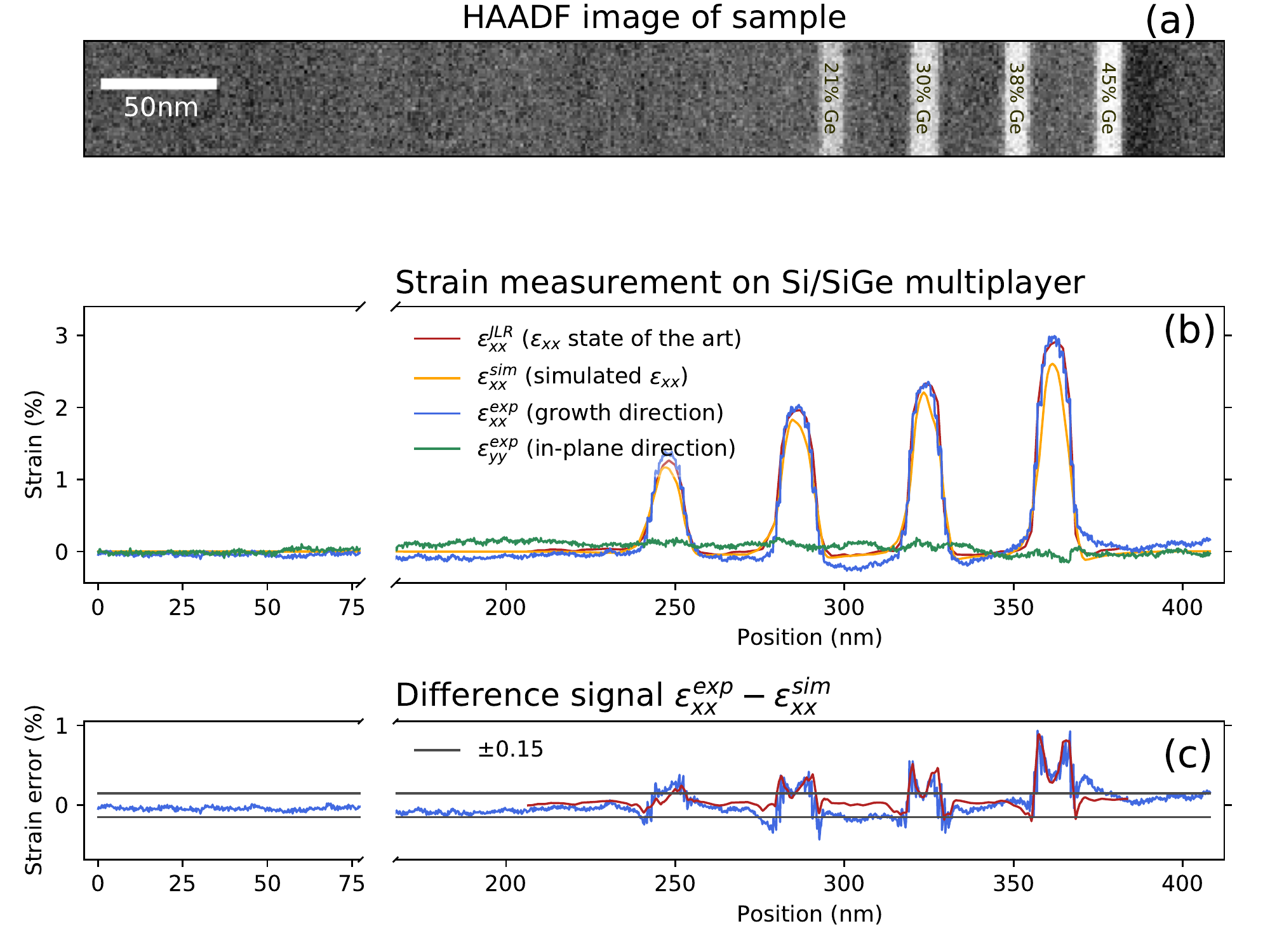}
    \caption{Experimental measure of strain. (a) Scanning transmission electron microscopy (STEM) image of the multilayer sample. (b) Strain recovered from the experimental data, compared with a finite element simulation of the strain profile, and with the N-PED measurement from Rouvière \textit{et al}. \cite{Rouviere2013}. (c) Difference between measured and simulated strain. The difference is below 1.5e-3, except at the interface where it is likely a matter of resolution/probe shape. The red curve, consisting of only 96 points, appears deceptively smoother. \label{fig:strainplot}}
\end{figure}

When applied to the simulated diffraction patterns, we get the results shown in figure \ref{fig:simulations}c. We can see immediately that the input strain is recovered with good accuracy ($ \Delta < 2 \times 10^{-3}$ ) and precision ($\sigma = 4.6 \times 10^{-4} $), though the value is slightly too low. The performance in the simulations appears  between those of NBED ($\sigma=11 \times 10^{-4}$) and N-PED ($\sigma=2 \times 10^{-4}$) \cite{Mahr2015}.

We implemented this idea experimentally by fabricating an annular aperture, which, placed in the illumination system of a TEM, realises the desired hollow-cone illumination. The illumination lenses of the TEM focus the electron beam on the sample, generating a fine electron probe with a wave function given by the aperture's Fourier transform. Alternatively, the aperture constitutes the momentum spectrum of the impinging beam, and hence each point of the ring in the aperture corresponds to a tilted impinging plane wave.
Such an aperture with a diameter of  $20 \, \mu m$ and an annulus width of $0.9 \, \mu m $ was manufactured by milling a $2\, \mu m$ thick gold film with the focused ion beam of a dual beam instrument, and inserted in a probe-corrected Titan$^3$ operated at $300\,kV$.
The semi-convergence angle $\alpha$ needs to be chosen carefully, as it impacts greatly the obtained result. If $\alpha$ is too low (far below the Bragg angle $\theta_B$ of the lowest -order excited reflection), the spatial resolution will be compromised. If $\alpha$ is much higher than $\theta_B$ the very large number of overlapping rings makes the patterns difficult to analyze.
We have found that values of $\alpha$ slightly below or equal to $2 \times \theta_B$ provide a good compromise between spatial resolution and overlap. In this case we have used $\alpha=5.9\,\textrm{mrad}$.
The semi-convergence angle is widely tunable on any microscope with an illumination system made of three condenser lenses and it can also be tuned, with a few limits, on other microscopes, though in the last case larger variations may require a further aperture replacement.

Since this microscope is not fitted with an image corrector, a direct observation of the probe shape might be unreliable, and the expected resolution is best estimated on theoretical grounds, then compared to the sharpness observed in the experimental features.
For a conical illumination from an infinitely thin annular aperture, the probe shape is described by a Bessel auxiliary function $J_0 (\alpha k_0 r_\perp )$, though in a realistic situation with a finite-width annulus this is only an approximation and the probe shape needs to be simulated numerically \cite{Grillo2016a,Saitoh2016a}. 

For conventional non-aberrated electron probes the wave function is described by an Airy disc, and the resolution is assessed through the full width at half maximum of its probability distribution. In previous experimental demonstrations of strain retrieval with NBED or N-PED, the resolution has been estimated at  $\approx 0.9\,nm$ \cite{Beche2009,Rouviere2013}.
While in the experimental conditions used ($\alpha = 5.9 mrad$) the simulated probe profile has a full width at half maximum of $70 \, pm$, that is not a good measure of resolution owing to the long (though weak) "tails" of the intensity distribution \cite{Grillo2014a,Grillo2016a,Saitoh2016a}. We therefore chose to estimate the resolution by comparing the sharpness of features between HAADF-STEM images acquired with a conventional high-resolution probe and with our modified beam (see Supplementary Material) obtaining a resolution of $1.3\,$nm. The width of these beams also depends heavily on the width of the annular slit, and a better resolution is likely possible (see Supplementary Material). These approximate electron Bessel beams are expected to be resistant to spherical aberrations \cite{Grillo2014a,Grillo2016a} and the performance demonstrated here could potentially be reproducible in older non aberration-corrected TEMs. While distortions in the projection systems can potentially affect the recorded pattern, their impact on the measured strain is negligible as long as this is measured relatively to a reference region \cite{Mahr2019}.

Using this setup, we test our proposed method on a well known and characterised sample: a stack of Si/SiGe layers grown on a silicon substrate \cite{Beche2009,Rouviere2013,Cooper2016}.
We scanned the beam in a linear fashion while using a conventional charge coupled device camera to record one diffraction pattern for each  beam position.
By using the sample morphology as observed from the microscopy images, as well as the concentration of Ge in the SiGe layers as measured by secondary ion mass spectroscopy \cite{Beche2011}, we also performed a finite element modeling (FEM) simulation that can be compared to the experimental data as shown in figure \ref{fig:strainplot}b.
For a more accurate assessment of resolution, the simulated strain has been convoluted with the simulated probe intensity.
By measuring the root-mean-square value of strain in the first part where no strain is expected, we can assess the noise level and hence the strain precision of the technique, to be better than  $\sigma = 2.5 \times 10^{-4}$.
The match with the simulated strain appears very accurate. Also shown is a strain measurement acquired by N-PED on a lamella of the same sample, and previously published by Rouvière \emph{et al.} \cite{Rouviere2013}. This measurement is used here to represent the state of the art.

The difference between the three plots appears small. Figure \ref{fig:strainplot}c shows that the match between experiment and simulations, assessed in the regions between the layers, is $1.5 \times 10^{-3}$ or better confirming the good accuracy, and appears close to the performance of N-PED.
The difference at the layer's edges is presumably due to imperfections in our hypothesized probe shape, while the the peak height isn't considered as a measure of accuracy as it likely suffers from imperfections in SIMS data used in the FEM modelling, and a similar deviation is observed in both N-PED and Bessel diffraction.

With this comparison we show how Bessel beam diffraction can detect strain with performances (precision $ 2.5\times 10^{-3} $, accuracy $ 1.5\times 10^{-3} $) which appear superior to those reported for NBED and approach those of N-PED. While Bessel beam diffraction can't cover many other use-cases supported by PED (e.g. orientation mapping or determination of complex crystal structures) it does not require expensive specialised hardware and can be potentially implemented in any current generation TEM with minimal instrumental modifications (and thus disruptions) and downtime. Furthermore, the technique requires no further alignment beyond a standard STEM experiment and the analysis code is freely available \cite{GhielensCode,DataDrop}, making it experimentally very accessible and an interesting approach for the study of strain.

\begin{acknowledgements}

GG acknowledges support from a postdoctoral fellowship grant from the Fonds Wetenschappelijk Onderzoek - Vlaanderen (FWO). AB and JV ackowledge support from the FWO under the project G.0934.17N ``Compressed sensing enabling low dose imaging in transmission electron microscopy''. CM acknowledges support from the Deutsche Forschungsgemeinschaft (DFG) under contract no. RO2057/12-2 within the research unit FOR2213.
We gratefully thank the CEA-LETI, Grenoble, for providing us with the sample.
\end{acknowledgements}

\nocite{*}
\bibliographystyle{apsrev4-1}
\bibliography{Bessel_Beam_Strain_Maps}

\section*{Supplementary information}

\subsection{Electron diffraction simulations:}
Multislice simulations have been performed with the STEMsim software \cite{Rosenauer2008}. To account for thermal diffuse scattering, we used the frozen-lattice approach \cite{VanDyck2009} averaging intensities of 15 diffraction patterns, simulated with different statistical, Gaussian distributed, displacement of the individual atoms according to the Debye-Waller factor at a temperature of 300 K \cite{Schowalter2009}. As inelastically scattered electrons are not taken into account, these simulations represent energy filtered diffraction patterns. Simulation parameters have been chosen to represent experimental conditions on an FEI Titan 80/300 TEM/STEM microscope operated at an acceleration voltage of $U = 300\,kV$ using a spherical aberration constant of $C_s = 1.2\,mm$. 
The simulated sample is a Si sample containing two embedded SiGe layers with Ge concentrations of 38\% and 31\%. The layer with higher Ge content shows a segregation profile according to the model of Muraki et al. \cite{Muraki1992} with a Ge segregation efficiency of R = 72\% \cite{Prestat2013}. The simulated crystal is a quadratic super cell with a width of 145 unit cells in the [001] growth direction and a specimen thickness of $50\, nm$. The sample is viewed along the [110]-zone axis using a beam semi-convergence angle of $5.9\,mrad$. We used an image of the experimental ring-shaped condenser aperture as C2 condenser aperture for the simulations.
In order to represent experimental diffraction patterns as accurately as possible, simulated diffraction patterns have been further modified according to Ref. \cite{Grieb2017}. These modifications include an additive background intensity, Poisson noise and a blurring caused by the modulation transfer function (MTF) of the CCD-camera used for the acquisition of experimental images. For the simulations evaluated for this report, we used the MTF of a Gatan Ultra-Scan 1000 CCD camera \cite{Krause2013}.

\subsection{Finite element modeling simulations:}
Finite element simulations were carried out
in the 2D plane strain approximation. In order to take into account the anisotropic behavior of crystalline silicon and silicon-germanium, the 2D compliance matrix was considered in the calculations. Such analysis allows to properly simulate the different mechanical behavior of the growth direction, following the [001] zone axis, from the in-plane plane direction, following [110] orientation \cite{Beche2011}.

\subsection{Data Analysis:}
The data analysis was performed with the code written ad-hoc in python/numpy, capable of parallel execution and highly efficient. On our computer it can analyse 2500 diffraction patterns in approximately 3 minutes. This code is available under GPLv3 License at the address:
\href{https://bitbucket.org/lutosensis/tem-thesis/src/master/}{https://bitbucket.org/lutosensis/tem-thesis/}

\subsection{Data Availability:}
The experimental data as well as the simulation results have been uploaded on the Zenodo repository at the address:
\href{http://dx.doi.org/10.5281/zenodo.2566137}{http://dx.doi.org/10.5281/zenodo.2566137}

\subsection{Intensity Profiles of Bessel and Airy probes:}
\includegraphics[width=\columnwidth]{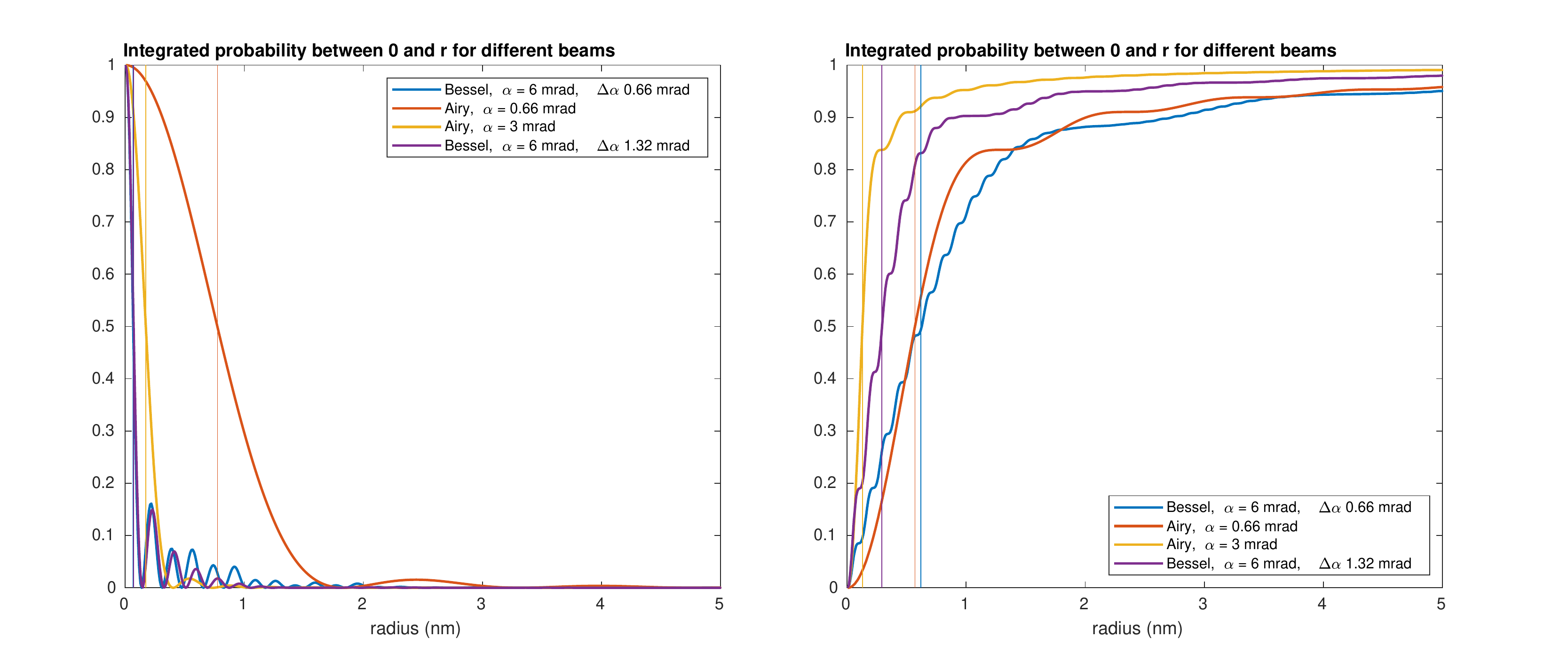} \\
Comparison of intensity profiles and cumulative integrated intensities (i.e. intensity contained within a given radius) for an approximate Bessel beam with a semi-convergence angle $\alpha=6\,$ mrad and a width of the annulus of 11\% of the radius, an Airy disc with $\alpha=0.66\,$ mrad (i.e. equal to the width of the annulus), a more conventional Airy disc with $\alpha = 3$ mrad, and a Bessel beam generated with a 22\% wide annulus. Notice the counterintuitive behaviour of Bessel beams here: despite having the sharpest central spot, they do have significant rippling which makes the radius of 50\% integrated intensity much larger than one would expect.

\subsection{Test of resolution by HAADF imaging}
\includegraphics[width=0.65\columnwidth]{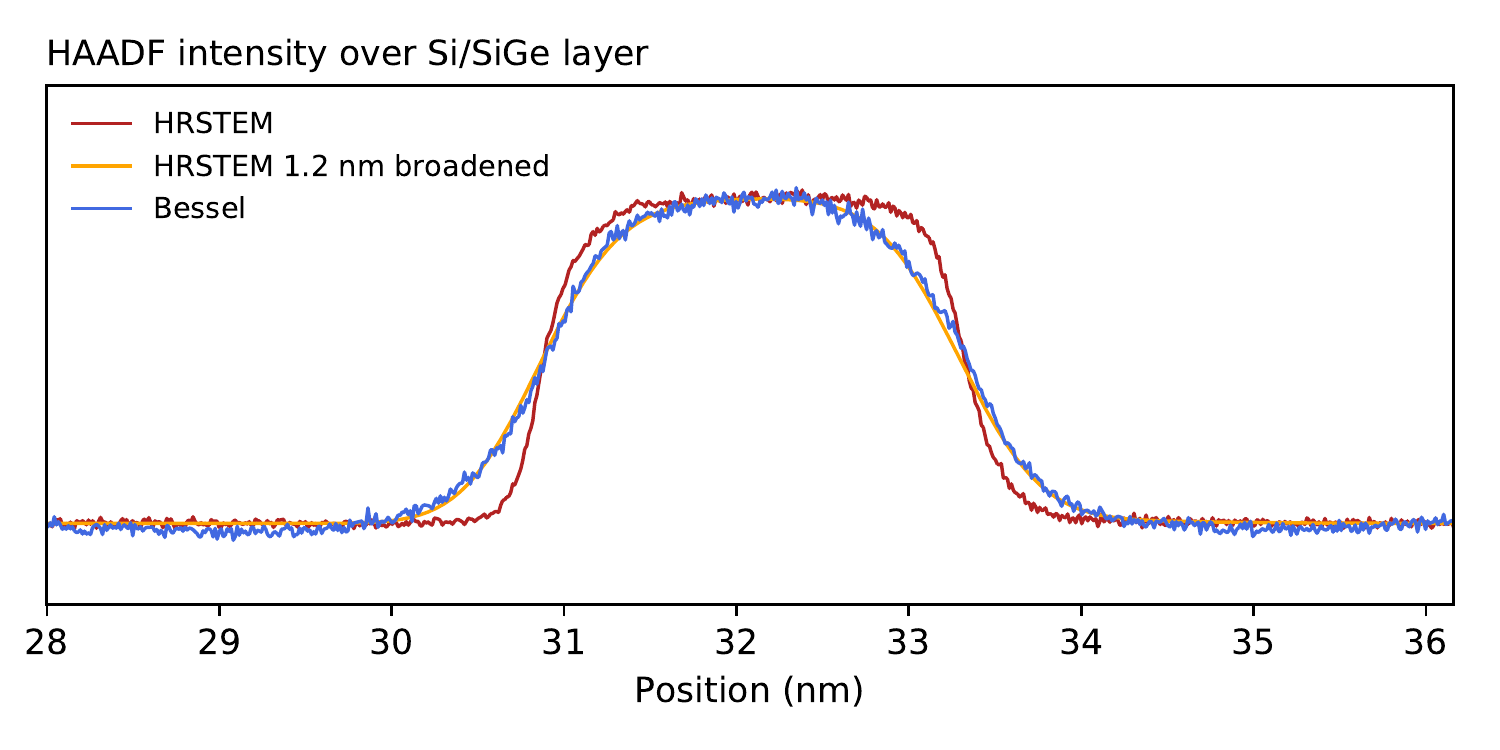} \\
Comparison of intensity profiles extracted from HAADF-STEM images. The sample used in the paper was imaged in HAADF configuration. Images were acquired in a conventional high resolution STEM configuration with sub-angstrom resolution, as well as with a Bessel beam with a convergence angle of about 5.9 mrad such as those used in the strain measurement. A line profile over one of the layers is shown for the conventional high resolution image (in red, the profile has been averaged in the lateral direction over several unit cells to improve signal), as well as the same profile convoluted with a Gaussian with a width of 1.2 nm (in orange), and an intensity profile extracted from the image acquire with the Bessel beam (in blue), The match between the orange and blue profiles suggest that the resolution achieved is better than 1.3 nm.


\end{document}